# Investigation into the potential use of Poly (vinyl alcohol)/Methylglyoxal fibres as antibacterial wound dressing components

Sophie E L Bulman[1] Parikshit Goswami[1] Giuseppe Tronci[1,2] Stephen J Russell[1] Chris Carr[1]

[1] Nonwovens Research Group, Centre for Technical Textiles, School of Design, University of Leeds, UK

[2] Biomaterials and Tissue Engineering Research Group, School of Dentistry, University of Leeds, UK

## Abstract

As problems of antibiotic resistance increase, a continuing need for effective bioactive wound dressings is anticipated for the treatment of infected chronic wounds. Naturally derived antibacterial agents, such as Manuka honey, consist of a mixture of compounds, more than one of which can influence antimicrobial potency. The non-peroxide bacteriostatic properties of Manuka honey have been previously linked to the presence of methylglyoxal (MGO). The incorporation of MGO as a functional antibacterial additive during fibre production was explored as a potential route for manufacturing wound dressing components. Synthetic MGO and polyvinyl alcohol (PVA) were fabricated into webs of sub-micron fibres by means of electrostatic spinning of an aqueous spinning solution. Composite fabrics were also produced by direct deposition of the PVA-MGO fibres onto a preformed spunbonded nonwoven substrate. Attenuated Total Reflectance Fourier Transform Infrared (ATR-FTIR) and Proton Nuclear Magnetic Resonance ($^1$H-NMR) spectroscopies confirmed the presence of MGO within the resulting fibre structure. The antibacterial activity of the fibres was studied using strains of Staphylococcus aureus and Escherichia coli. Strong antibacterial activity, as well as diffusion of MGO from the fibres was observed at a concentration of 1.55mg/cm².

## Introduction

Chronic wounds such as pressure ulcers and leg ulcers, cause patients pain and delayed healing occurs as a result of infection and biofilm formation on the surface of the wound[1]. Bacteria biofilms can harbour a host of different bacterial species including aerobic and anaerobic microorganisms, some of which include Methicillin-resistant Staphylococcus aureus (MRSA), Escherichia coli (E. Coli), beta-haemolytic Streptococci and Enterobacter cloacae[2]. As these bacteria start to show resistance to antibiotics[3, 4], the use of topical wound dressings, containing antimicrobial compounds are an important tool for the treatment of infected chronic wounds. Numerous antimicrobial compounds used in combination with dressings have been clinically evaluated. Of these, the use of silver in topical antibacterial wound dressings[5, 6] is particularly well established although potential toxic effects, such as argyria[7] and cytotoxicity, have been reported in wound healing[8, 9]. Naturally occurring antimicrobial agents include Manuka honey, which relies upon an osmotic effect produced by the high sugar content, the presence of an enzyme that produces hydrogen peroxide and non-peroxide compounds to provide antibacterial function[10]. Methylglyoxal (MGO)[11], a ketoaldehyde (Figure 1), is a non-peroxide antimicrobial compound present in Manuka honey and is a metabolite found within the human body and many other living organisms[12].

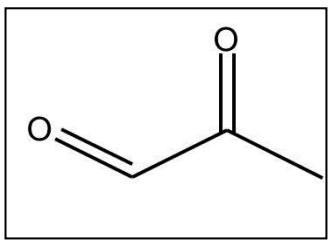

**Figure 1**. Chemical structure of Methylglyoxal

MGO exhibits strong activity against malignant cancer cells and can stimulate the immune response system to target tumour cells[13]. Antiviral[14] and antimalarial properties have also been reported[15], as well as its use in liquid form as an antibacterial agent for wounds[16]. The toxicity of MGO in the presence of bacteria is thought to be attributed to MGO's ability to modify cell compounds, including free amino acids[17, 18], proteins [17, 19, 20] and nucleic acids[21]. It has also been reported that MGO is able to modify DNA in leukaemia cells and prostate cancer cells by the induction of apoptosis[13, 22]. Another study reported that MGO was able to activate lymphocytes and macrophages against tumour cells[23]. Lymphocytes and macrophages also play an important role in the wound healing process[24]. MGO presents potential complications for patients with diabetic foot ulcers due to the formation of irreversible advanced glycation end products (AGEs) which can change collagen pathophysiology resulting in the disruption of normal collagen matrix remodelling [25]. Another study has reported that MGO may lead to vascular complications in patients with diabetes[26].

Many clinically utilised wound dressings containing antimicrobials are based on nonwoven substrates, which are porous three-dimensional fibre assemblies. These dressings provide a vehicle to deliver the antimicrobial agent, but this is in addition to providing many other major functions critical to wound healing such as wound exudate management. Fibre-forming hygroscopic materials and hydrogels are particularly important raw materials in wound dressing design.

Polyvinyl alcohol (PVA) is a polyhydroxy polymer, which is known to have good fibre forming properties[27]. It is water-soluble, biocompatible and approved by the FDA for medical use in humans[28]. PVA is also valuable as a hydrogel forming polymer. Hydrogels have a distinct advantage in wound healing, since they are water absorbing gel substances of varying rigidity[29] that are able to provide a well-defined moist wound environment that allows a wound to heal faster than in a dry state[30]. The moist environment promotes epithelial cell migration from the wound edges, encourages modification of pH and oxygen levels, maintains an electrical gradient and retains wound fluid on the wound surface[31]. PVA has been previously studied as a potential wound dressing material because of its hydrogel forming properties[32, 33] and its ability to provide controlled release of antibiotics into rats[34]. A recent study discussed PVA based hydrogel fluids, which on contact with glucose present in wound exudate formed a gel that moulded to the shape of the wound[35]. Typically, in the manufacture of nonwoven dressings, the antimicrobial compound is impregnated in to the nonwoven such that the constituent fibre surfaces are topically coated or it is incorporated within the fibres.

Electrostatic spinning of PVA fibres containing a limited range of antibacterial agents for wound healing has previously been reported[36, 37] and could provide a potential route for manufacturing thin fibre web layers that can be incorporated in to composite dressings. Additionally, the high specific surface area and porosity of electrospun fibre webs have the potential to absorb large volumes of fluid exudate from the wound, inhibiting exogenous microorganisms from entering the wound and to aid in fluid drainage[38]. The production and analysis of PVA-synthetic MGO fibres has yet to be systematically studied. The purpose of this study was to determine the feasibility of manufacturing fibres and fabrics from mixtures of PVA and MGO and to produce antibacterial fibres for potential use as part of a composite wound dressing material.

**Experimental**

**Materials and methods:**

*Materials:*

Poly (vinyl alcohol) with a molecular weight of 31-50,000 and a 40 wt% methylglyoxal aqueous solution were purchased from Sigma Aldrich UK. Polypropylene spunbond fabric (SPB) with a weight of 0.5g/m² was purchased from Elmarco s.r. in the Czech Republic.

*Preparation of PVA and PVA/MGO spinning solutions:*

Two separate solutions were prepared in the absence or presence of MGO. Preliminary experimental work indicated that an 8% (w/v) PVA solution enabled stable electrospinning conditions. In the former case, 0.8 g of PVA was dissolved in 10 ml of distilled water (Solution 1). For the preparation of MGO-containing solutions, 10 ml of 40 wt%

MGO solution was diluted with 30 ml of distilled water to achieve a concentration of 11.22% (prepared MGO solution). 0.8g of PVA was dissolved in 10 ml of the prepared MGO solution (Solution 2). Both solutions were agitated for 2 hr at 80°C by magnetic stirring and left to cool to room temperature.

**Characterisation of PVA and PVA/MGO spinning solutions**

Viscosity measurements were made using a Brookfield DV-E Viscometer. Readings were taken at $22^0$C using spindle S18 at 60 r min$^{-1}$. The surface tension of the solutions was measured using a Kruss tensiometer K100. A sample vessel with a diameter of 66.5 mm and a platinum plate was used to take the measurements at $22^0$C.

*Preparation of PVA and PVA/MGO fibre webs:*

The electrospinning equipment consisted of a syringe pump (KD Scientific Model 200 Series), a Glassman high voltage power supply and a grounded square (10 x 10cm) of polypropylene spunbond (SPB) nonwoven fabric used as the collector, was placed over a square (10 x 10cm) piece of aluminium foil. This enabled fibre webs to be directly combined on to the surface of a pre-formed reinforcing nonwoven fabric, to produce a composite. A 10 mL syringe was fitted with a 21 gauge blunt needle and the polymer solution was loaded into the syringe which was set to a feed rate of 0.1 mL hr$^{-1}$ for 5 hr with a needle tip to collector distance of 10 cm and a voltage of 10kV. The temperature and humidity of the chamber was $20^oC \pm 2^oC$ and 43% ± 2%, respectively.

**Characterisation of PVA/MGO fibres**

*Fibre Morphology*

The as-spun fibres in each collected fibre web were inspected using a Field Emission Gun Scanning Electron Microscope (FEGSEM) (Carl Zeiss LEO 1530). The mean fibre diameter was determined based on evaluation of individual fibres in the SEM images; the total number of measurements for each sample was 50.

*Fourier-transform infra-red (FTIR) spectroscopy*

Infra-red spectra of the PVA powder, 40 wt% MGO solution and PVA/MGO fibres were obtained using an FT-IR Perkin Elmer Spectrum BX with diamond ATR attachment system. The PVA powder and 40 wt% MGO solution were analysed with the diamond ATR system. PVA/MGO fibres were cut into small pieces and ground with KBr powder (Sigma Aldrich Chemicals) to make sample disks. This was done to ensure any MGO encapsulated within the fibres was made available for detection. Measurements were taken in the range between 4000 – 400 cm$^{-1}$ with a resolution of

4 cm$^{-1}$. 64 repeat scans were averaged for each spectrum of the PVA powder and 40 wt % MGO solution, respectively, and 16 repeat scans were averaged for each spectrum of the PVA/MGO fibre KBr disk.

*$^1$H-NMR Spectroscopy*

10 mg of PVA powder, 40 wt% MGO solution and PVA/MGO fibres, respectively, were dissolved in 1mL of deuterium oxide and analysed in a Bruker Advance spectrometer 500 MHz $^1$H NMR spectrometer. The $^1$H-NMR spectra were the average of 1024 repetitions.

*Antibacterial studies:*

The PVA fibre SPB fabric composites, PVA/MGO fibre SPB fabric composites and SPB (polypropylene spunbond nonwoven fabric) alone were tested for antibacterial activity according to a standard protocol, BS EN ISO 20645:2004 – Textile fabrics, determination of antibacterial activity, agar diffusion plate test. Both gram-positive and gram-negative strains of bacterium were used, which included Staphylococcus aureus and Escherichia.coli. Both of the strains were used in accordance with the requisite standards and are common pathogens found in infected wounds. After 24 hours incubation at 37ºC, the microbial zone of inhibition was measured using optical microscopy.

**Results and discussion**

*Fibre morphology*

Figures 2a and 2b show the morphology of fibres in fibre webs with and without MGO, respectively. Both reveal evidence of bead formation in the as-spun fibres, but the frequency was reduced in the fibres produced from the PVA-MGO solution (Figure 2b). Beads were formed during electrospinning due to the capillary breakup of the electrospinning jets as a result of surface tension, which is affected by the presence of electrical forces[39]. Viscosity, which is strongly influenced by the concentration of the solution, is also key parameter pertinent to the elimination of beads[40]. The viscosity of the MGO-containing spinning solution was almost twice as high as the solution containing only PVA (Table 1). The increased viscosity affected the fibre morphology as evident in Figure 2b, where there appeared to be fewer beads along the length of individual fibres. By altering the solution concentration it is possible to increase the viscosity and the surface tension, thus smoothing the fibre morphology and reducing the formation of beads[41]. By reducing the molecular weight of the PVA, there will be fewer chain entanglements within the solution and a higher amount of free solvent. The free solvent molecules will tend to accumulate and form spherical beads due to the surface tension of the solution. Previous studies have also shown that addition of salts can increase the net density charge of the solution and by the addition of ionic salts, the charge density can be controlled in order to decrease the amount of bead formation[42, 43]. Temperature and humidity also affect fibre formation because of their influence on the evaporation rate of the solvent, which changes the viscosity of the polymer solution[44], but in the present work these effects were minimised by maintaining constant temperature and humidity in the spinning chamber.

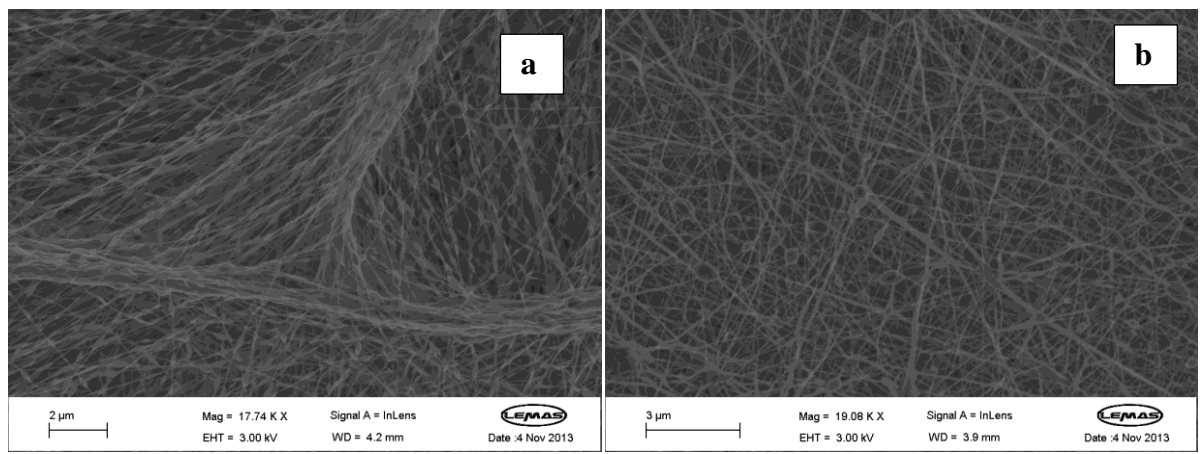

**Figure 2**. SEM micrographs of as-spun PVA fibres, (a), mean fibre diameter = 118nm and PVA/MGO fibres, (b), mean fibre diameter = 166nm.

**Table 1.** Properties of aqueous PVA spinning solutions.

| Solution | Composition of solution | Surface Tension (mN/m) | Viscosity (cPs) |
|---|---|---|---|
| 1 | 8% (w/v) PVA | 46.5 | 18.5 |
| 2 | 8% PVA (w/v) in prepared MGO solution | 52.2 | 35.7 |

*Fourier-transform infra-red (FTIR) spectroscopy*

To determine the presence of MGO within the as-spun fibres, FTIR analysis was conducted in order to identify the absorption bands of MGO within the PVA/MGO fibres, Figure 3. MGO is a ketoaldehyde with a characteristic vibrational absorption peak at 1720 cm$^{-1}$, which is attributed to the C=O stretching vibration in the ketone carbonyl. A peak at 1379 cm$^{-1}$ was also observed and can be assigned to $CH_3$ bending in MGO[45]. These characteristic spectral vibrations confirm the presence of MGO within the fibres. The other vibrational peaks observed in the spectra can be assigned to the base PVA polymer where the broad band centred at 3282 cm$^{-1}$ is due to hydrogen-bonded O-H stretching vibration and the peak at 2905 cm$^{-1}$ is attributed to C-H stretching in the polymer backbone. The peak at 1419 cm$^{-1}$ is the result of in-plane O-H bending. The peak at 1089 cm$^{-1}$ is indicative of C-O stretching[45].

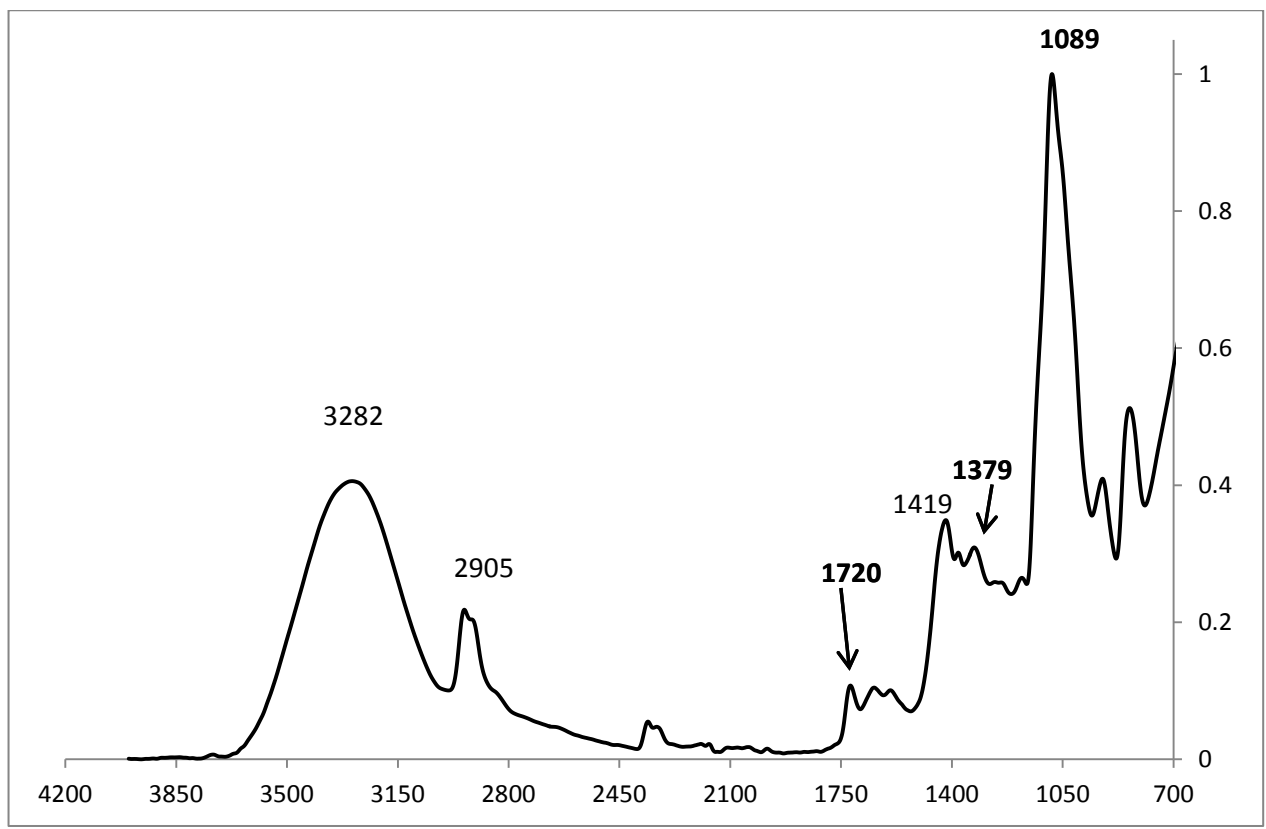

**Figure 3**. ATR-FTIR spectrum of PVA/MGO fibres

*$^1$H-NMR Spectroscopy*

In order to further verify the presence of MGO within the PVA/MGO fibres, $^1$H-NMR spectra of fibres obtained from MGO-containing solutions were recorded. The $^1$H-NMR spectra seen in Figure 4 of PVA/MGO fibres showed two singlet resonances at 1.408 and 2.337 ppm which can be assigned to the methyl protons present in the methylglyoxal di-hydrate and monohydrate, respectively, following the reaction of MGO with water[46]. In addition the resonance peak at 5.309 ppm is assigned to be the alkyl proton of methylgyoxal monohydrate, which is in agreement with previous studies[46]. The resonance peaks observed at 1.750 ppm and 4.075 ppm can also be assigned to the protons of $CH_2$[47] and the CH proton of the PVA, respectively. These spectral assignments have further confirmed the presence of MGO within the fibres. Consequently, solvent evaporation during electrospinning has proved to have minimal impact on MGO loading of resulting PVA fibres.

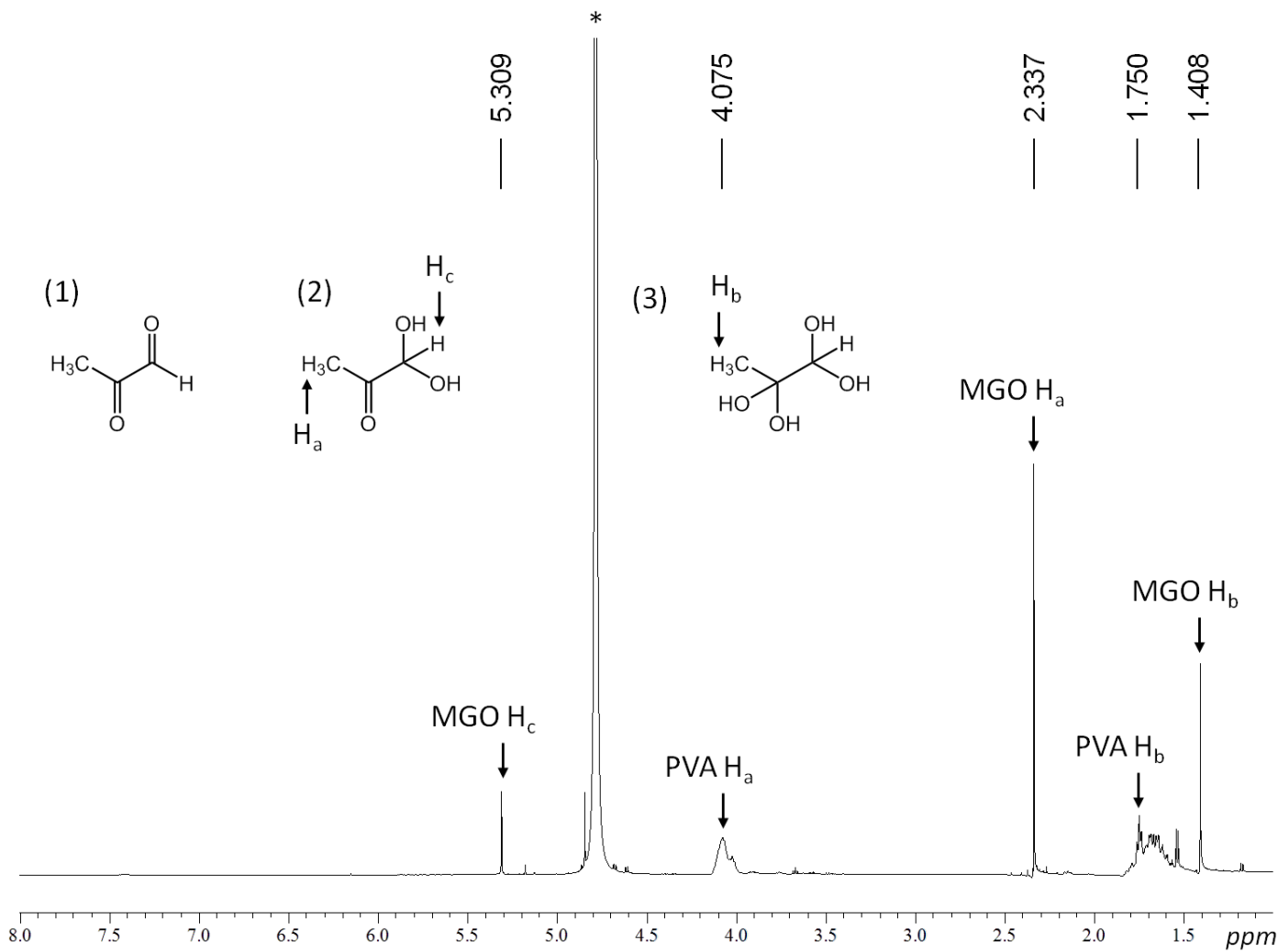

**Figure 4.** ¹H-NMR spectra of PVA/MGO fibres. (1), (2) and (3) indicate the chemical formulas of MGO, MGO mono-hydrate and MGO di-hydrate, respectively.

*Antibacterial Studies*

Figure 5 shows visual images of the effect of fabric samples on the growth of E. coli and S. aureus after 24 hours incubation. Table 2 indicates the zone of inhibition for all samples and the calculated percentage concentrations of MGO in the PVA fibres. In the agar diffusion plate tests zones of inhibition were clearly evident only in the PVA/MGO SPB composites after 24 hours incubation, with the PVA/SPB and SPB samples functioning as experimental controls. The contact zone underneath the samples was examined for bacterial growth using optical microscopy at 20X magnification. There was no evidence of bacterial growth below the PVA/MGO samples, while moderate growth was evident below the PVA SPB and SPB samples. Thus, in the absence of MGO, there was insufficient antibacterial functionality to restrict growth. These results confirm that the fibre webs containing MGO exhibit toxicity against both common gram positive and gram negative strains of bacterium, with zones of inhibition of 9.1 mm (E. coli) with a MGO concentration of 1.55 mg cm$^{-2}$ and 11.4 mm (S. aureus) with a MGO concentration of 2.35 mg cm$^{-2}$.

A previous study reported the bactericidal activity of MGO in liquid and gel formulations against Staphylococcal species, including S. aureus and methicillin-resistant S. epidermidis (MRSE)[16]. Concentrations between 1.25 mg mL$^{-1}$ (0.14 mg cm$^{-2}$) and 15 mg mL$^{-1}$ (1.72 mg cm$^{-2}$) were tested in a 9 mm well. The results showed that at the lower concentration of 0.14 mg cm$^{-2}$, a zone of inhibition of approximately 20 mm for S. aureus and 16 mm for MRSE was formed. At the higher concentration of 1.72 mg cm$^{-2}$ zones of approximately 40 mm were recorded for both strains.

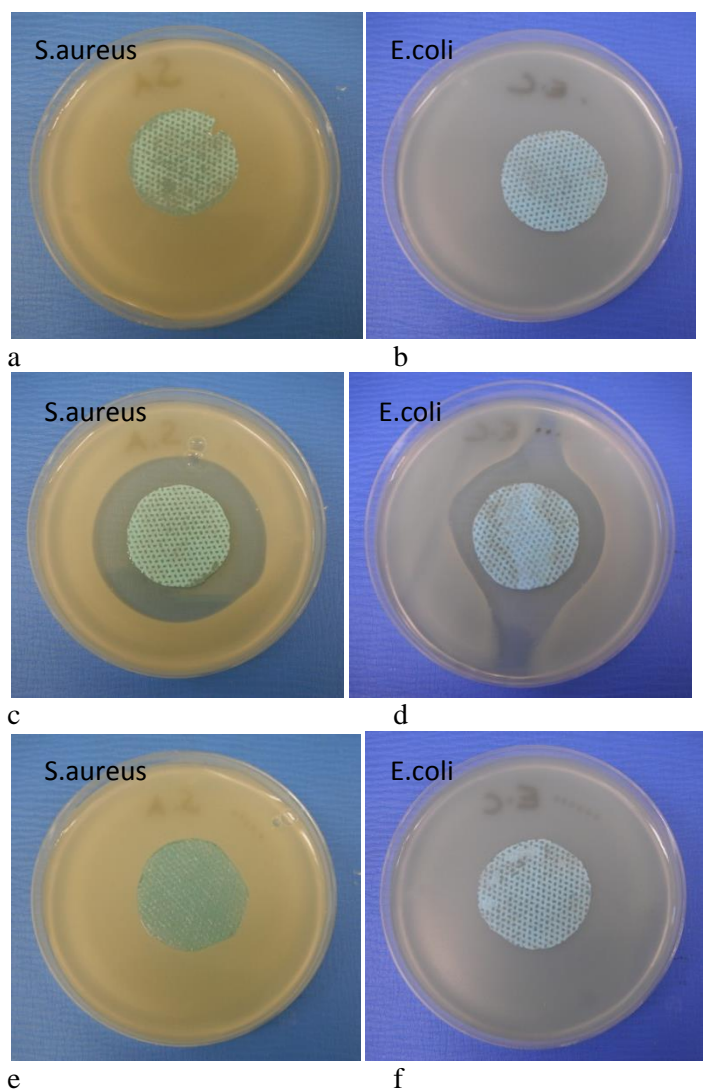

**Figure 5.** Antibacterial activity of PVA fibres on a spunbond nonwoven substrate, (a) and (b) PVA/MGO fibres on a spunbond substrate, (c) and (d) and base spunbond nonwoven substrate, (e) and (f).

These results highlight the fact that when MGO is in liquid or gel form, with a concentration of approximately one tenth (0.14 mg cm$^{-2}$) than that present in the PVA/MGO fibre webs (1.55 mg cm$^{-2}$) a zone of inhibition twice the size can be achieved. This difference is not surprising given that the MGO is initially encapsulated within the fibre and is not initially freely available as in the liquid phase. On contact with the agar, the PVA fibres formed a gel, which promoted the diffusion of MGO from the fibres in to the surrounding bacterial strain, preventing bacterial growth. The kinetics of MGO release from the PVA fibres is the subject of further study and is likely to be affected by subsequent crosslinking. It is possible that some of the MGO may be retained in the PVA fibres after 24 hours depending on the molecular architecture of the hydrogel network.

**Table 2.** Zone of inhibition for PVA fibres on a spunbond nonwoven substrate, PVA/MGO fibre on a spunbond nonwoven substrate and base spunbond nonwoven substrate against S. aureus and E. coli.

| Sample ID | Sample | Bacteria Strain | Zone of Inhibition (mm) | Concentration of MGO (mg cm$^{-2}$) |
|---|---|---|---|---|
| a | PVA SPB | S. aureus | 0 | n/a |
| b | PVA SPB | E. coli | 0 | n/a |
| c | PVA/MGO SPB | S. aureus | 11.4 | 2.35 |
| d | PVA/MGO SPB | E. coli | 9.1 | 1.55 |
| e | SPB | S. aureus | 0 | n/a |
| f | SPB | E. coli | 0 | n/a |

**Conclusions**

PVA/MGO fibres were successfully prepared by the electrospinning technique and were directly deposited as a thin fibre web layer onto a preformed PP nonwoven substrate. Although the fibre morphology can be further optimised to reduce bead content, it was demonstrated that it is feasible to manufacture a mechanically robust composite structure comprising an antibacterial layer of hydrogel-forming polymer and MGO. The presence of MGO within the fibres was confirmed using both FTIR and NMR analyses. The antibacterial study revealed that MGO incorporated within PVA fibres, supported on a PP spunbond fabric, produces zones of inhibition for both gram positive and gram negative strains of bacterium, based on MGO concentrations in the fibre web of 2.35 mg cm$^{-2}$ for S. aureus and 1.55 mg cm$^{-2}$ for E. coli.

**Acknowledgements**

The authors would like to thank Robert Davies for $^1$H-NMR spectra acquisition, Peter Parnell for assistance with the antimicrobial studies.

**Funding**

The financial support of the Clothworkers' Foundation and the Clothworkers' Centre for Textile Materials Innovation in Healthcare is gratefully acknowledged.